\documentstyle[12pt]{article}
\title{\Large Non-Local Pseudo-Differential Operators}
\author{D.\ G.\ Barci\thanks{barci@vmesa.uerj.br} \\ Universidade do Estado do Rio de Janeiro,\\
Instituto de F\'\i sica, Departamento de F\'\i sica Te\'orica \\
R.\ S\~ao Francisco Xavier, 524,\\ Maracan\~a, cep 20550, Rio de Janeiro, Brazil.       
\and 
C.\ G.\ Bollini \\  Universidad Nacional de La Plata,\\
C.C. 67 (1900), La Plata, Argentina
\and
L.\ E.\ Oxman\thanks{oxman@if.ufrj.br} \\Instituto de F\'\i sica, Departameto de
F\'\i sica Te\'orica\\
Universidade Federal do  Rio de Janeiro \\
C.P. 68528, Rio de Janeiro, RJ, 21945-970, Brazil. 
\and
M.\ C.\  Rocca \thanks{rocca@venus.fisica.unlp.edu.ar}\\Universidad Nacional de La Plata, \\
C.\ C.\ 67 (1900), La Plata,  Argentina. \\
Universidad Nacional del Centro.\\
Pinto 399, Tandil, Pcia.\ de Bs.\ As.\, Argentina}
\date{April 10, 1996}
\begin{document}
\maketitle
\newpage
\begin{abstract}
We define, in a consistent way, non-local pseudo-differential operators acting on a space of analytic functionals. These operators include the fractional
derivative case. In this context we show how to solve homogeneous and
inhomogeneous equations associated with these operators.
We also extend the formalism to d-dimensional space-time solving, in particular,
the fractional Wave and Klein-Gordon equations.  
\end{abstract}
\newpage
\section{Introduction \label{s1}}

Interest in non-local field theories has always been present in theoretical
physics, associated to several different motivations. 

In Ref.\ \cite{1}, J.\ A.\ Wheeler and R.\ P.\ Feynman considered a
description of the interaction between charged particles, where the
electromagnetic field does not appear as a dynamical variable (action
at distance). 

More recently, efforts were made to use non-local theories in connection
with the understanding of quark confinement and anomalies
\cite{3}\cite{4} and in string theories containing non-local vertices
\cite{5}\cite{6}.

Besides the possibility of non-local interactions, a field theory can also display non-local kinetic terms.

Before renormalization theory became well established the possibility
was considered of formulating finite theories by means of non-local 
kinetic lagrangians. A.\ Pais and G.\ E.\ Uhlembeck \cite{2} were one 
of the first to analyze non-local theories in this context.

The analytic regularization 
method introduced in Ref. \cite{7}, can be thought of as associated to
non-local kinetic terms in the lagrangian, specifically to fractional
wave and Klein-Gordon equations.

This type of non-locality also arises in effective theories, when integrating over some degrees of freedom in an underlying local field theory
\cite{8}\cite{9}\cite{10}. 

At a classical level, non-local equations containing arbitrary powers of
the D'alambertian ($\Box$)  have been studied in Ref. \cite{11}. In this reference  a non-trivial relation  between the number of
dimensions and the power of the operator was established  to satisfy the
Huygens principle (see Ref.\ \cite{12}).  In particular, in $2+1$
dimensions the usual wave equation leads to a Green function that
does not satisfy Huygens principle, while a non-local  equation with
$\Box^{1/2}$, does satisfy this principle (see also Ref.\ \cite{13}).
It is not by chance that the pseudo-differential operator $\Box^{1/2}$
appears also in the process of bozonization in $2+1$ dimensions.
In Refs.\ \cite{14} and \cite{14b} a mapping was established between Dirac's kinetic term and bosonic terms involving $\Box^{1/2}$. In Ref.\
\cite{15} it is shown how similar terms appear when $3+1$ dimensional
QED is projected to a physical plane. The kinetic term contains
$\Box^{-1/2}$, which in the static limit reproduces correctly the
$r^{-1}$ Coulomb potential, instead of the usual logarithmic behavior
of $2+1$ QED. This fact was first noticed in Ref.\ \cite{12}.

Also, in Ref. \cite{ffp}, a fractional generalized Fokker-Planck-Kolmogorov equation is proposed to describe anomalous transport in Hamiltonian systems, 
and in Refs.\  \cite{fd1}\cite{fd2} some particular Green functions for fractional diffusion and fractional wave equations were obtained.

In Refs. \cite{13} and \cite{20}, solutions to some particular non-local homogeneous equations were proposed on physical grounds. 

In this paper, we present a general mathematical approach to deal with pseudo-differential operators which enables us to obtain the whole space 
of solutions to homogeneous equations associated to them. Therefore, for a given pseudo-differential operator, it is possible to define a general Green function (having a general prescription to avoid the singularities).

In section \ref{s2}, we define the functional
spaces in which we build-up the framework for a correct description of
pseudo-differential operators. In these spaces we introduce the
representation of analytic functionals as
``ultra dis\-tri\-butions''\cite{16}\cite{17}, a convenient way to handle
and operate with the usual Green functions of Quantum Field Theories.

In Section \ref{s3} we define local and non-local operators acting on
ultra distributions. We write and solve formally, homogeneous and
inhomogeneous equations associated to them.

In section \ref{s4}, some examples are examined, such as rational and
power functions of the derivative operator. Solutions of the
respective homogeneous equations are given. Also, Green functions are
found, and their relations to fractional derivatives of Ref.\ \cite{8} are exhibited.

In section \ref{s5}, we extend our developments to space-time, by
defining non-local  functions of $\Box$. Also, fractional wave and
Klein-Gordon equations are introduced, and solutions for the respective
homogeneous equations are given. Corresponding Green functions are
found.

Finally, section \ref{s6} is devoted to a discussion of the
developments and results of the paper. 

\section{Functional Spaces \label{s2}}

We will start with the space $\zeta$ of entire analytic test
functions $\varphi(k)$, rapidly decreasing in any horizontal band.
We will call ``ultra-analytic'', any function $\varphi(k)\in \zeta$.
They are Fourier transforms of the space $\hat{\zeta}$ of all $C^\infty$
functions  $\varphi(x)$, such that $\exp(q|x|D^p\varphi(x))$ is
bounded in $R$ for any $q$ and $p$. 

In view of the latter property, $\zeta\supset Z$, where $Z$ is the
space of Fourier transforms of $K$ ($C^\infty$ functions on a compact
 set).  (\cite{18}, ch.\ 2, $\S 1.1$)

The dual of $\zeta$ is the space $\zeta'$ of linear functionals
defined on $\zeta$. In $\zeta'$ we can represent the propagators of a 
quantum field theory as analytic functionals, with the physical
properties that are expected from them \cite{19}. 

The general form of an analytic functional is: 
\begin{equation}
\psi(\varphi)=\int_L    \psi(k) \varphi(k)~dk
\label{e1}
\end{equation}

The ``density'' $\psi(k)$ is an analytic function and the line $L$ can
be  deformed as long as no singularity of $\psi(k)$ is crossed. $L$ can be an open line or a closed loop.

Not only $L$ can be deformed without altering $\psi(\varphi)$. Also,
the structure of singularities of the density can be modified. For
example,  when $\psi(k)=(k-\tau)^{-1}$, the development:
$(k-\tau)^{-1}=\sum_n \tau^n k^{-n-1}$, allows the pole at $k=\tau$
to be  represented by a series of multipoles located at the origin.
This fact is closely related to the expansion of the analytic
functional $\delta(k-\tau)$ as a series of $\delta^{(n)}(k)$
(\cite{18}, ch.\ 2, $\S 1.4$ ).  We can see then, that an analytic
functional can be expressed in more than one way. 

We are going to use, systematically, the following representation
\cite{16}\cite{17}:
\begin{equation}
\psi(\varphi)=\int_\Gamma dk~\psi(k) \varphi(k)~~~(\psi\in\zeta'~,
~\varphi\in\zeta)
\label{e2}
\end{equation}
where $\psi(k)$ is analytic in $\{k\in C:|Im\;k|>\rho\}$
and $\psi(k)/k^{\rho}$ is bounded continuous in $\{k\in C:|Im\;k|
\ge \rho\}$, $\rho$ depending on $\psi(k)$, $\rho\in N$
($N=$ set of entire numbers). 

The path $\Gamma$ runs from $-\infty$ to $+\infty$ along $Im\;k>\rho$
and from $+\infty$ to $-\infty$ along $Im\;k<-\rho$. 

If $\psi(k)$ is an entire function $a(k)$, then (\ref{e2}) is the zero
functional ($a(k)$ need not be ultra analytic). 
\begin{equation}
\psi_0=\int_\Gamma dk~a(k)\varphi(k)\equiv 0
\label{e3}
\end{equation}
Equation (\ref{e3}) tell us that $\psi(k)$ and $\psi(k)+a(k)$
represents one and the same functional $\psi(\varphi)$. 

If we take
\begin{equation}
\psi(k)=\frac{1}{2}Sg(Im\;k)
\label{e4}
\end{equation}
then
\begin{eqnarray}
\psi(\varphi)&=&\frac{1}{2}\int_\Gamma dk~Sg(Im\;k)\varphi(k)
 \nonumber \\
&=&\frac{1}{2}\int_{-\infty}^{+\infty} dk~\varphi(k)-
\frac{1}{2}\int_{+\infty}^{-\infty} dk~\varphi(k)   \nonumber\\
&=&\int_{-\infty}^{+\infty} dk~\varphi(k)=1(\varphi) \label{e5}
\end{eqnarray}
The density given by (\ref{e4}) implies that $\psi(\varphi)$ is the
unit functional. 

When we take $\psi(k)\equiv 0$ for $Im\;k>\rho$ (resp. for
$Im\;k<-\rho$) the ultra distribution is the retarded
(resp. advanced) functional. 

Feynman's causal function can also be represented as as
ultra distribution. In fact, the propagator is: 
\begin{equation}
\frac{1}{k_0^2-\bar{k}^2-m^2}=\frac{1}{k_0^2-\omega^2}=
\frac{1}{2\omega}\left( \frac{1}{k_0-\omega}-\frac{1}{k_0+\omega}
\right), ~~~ \omega>0
\label{e6}
\end{equation}
The causal Green function is such that the positive energy pole
(at $k_0=\omega$) is propagated by a retarded potential and the
negative energy pole propagates in an advanced way. To represent this
Green function we take
\begin{eqnarray}
\psi_F(k)&=&\frac{1}{2\omega}\frac{1}{k-\omega}
\mbox{~~,~~for~~}Im\;k>0 \nonumber\\
\psi_F(k)&=&\frac{1}{2\omega}\frac{1}{k+\omega}
\mbox{~~,~~for~~}Im\;k<0
\label{e7}
\end{eqnarray}
If we interchange the poles, taking into account the senses of the path
$\Gamma$, we get the  anticausal Green function: 
\begin{eqnarray}
\psi_A(k)&=&-\frac{1}{2\omega}\frac{1}{k+\omega}
\mbox{~~,~~for~~}Im\;k>0 \nonumber\\
\psi_A (k)&=&-\frac{1}{2\omega}\frac{1}{k-\omega}
\mbox{~~,~~for~~}Im\;k<0
\label{e8}
\end{eqnarray}
Wheeler propagator (half causal and half anticausal) is then

\begin{eqnarray}
\psi_W(k)&=&~~\frac{1}{2}\frac{1}{k^2-\omega^2}
\mbox{~~,~~for~~}Im\;k>0 \nonumber\\
\psi_W(k)&=&-\frac{1}{2}\frac{1}{k^2-\omega^2}
\mbox{~~,~~for~~}Im\;k<0\nonumber \\
\psi_W(k)&=&\frac{1}{2}\frac{Sg(Im\;k)}{k^2-\omega^2}
\label{e9}
\end{eqnarray}

We shall also need the Fourier transforms of the spaces $\zeta$ and
$\zeta'$. For a function 
$\varphi\in\zeta$, the Fourier  transform is: 
\begin{equation}
{\cal F}\varphi=\hat{\varphi}(x)=\int_{-\infty}^{+\infty}
dk~\varphi(k)e^{-ikx}
\label{e10}
\end{equation}
All functions $\hat{\varphi}(x)$ are $C^\infty$
and $\exp{q|x|D^p\hat{\varphi}(x)}$ is bounded for any $q$ and $p$.
They form the space  $\hat{\zeta}$. 

A natural definition for the Fourier  transforms of the functionals
$\psi\in \zeta'$ is:
\begin{equation}
{\cal F}\psi({\cal F}\varphi)=\hat\psi(\hat\varphi)
=2\pi\psi(\varphi)
\label{e11}
\end{equation}

For an ultra distribution represented by (\ref{e2}):
\begin{eqnarray}
2\pi\psi(\varphi)=\hat\psi(\hat\varphi) &=&
\int_\Gamma dk~\psi(k)\int_{-\infty}^{+\infty}dx~\hat\varphi(x)
e^{ikx} \nonumber \\
\hat\psi(\hat\varphi) &=&\int_{-\infty}^{+\infty}dx~\hat\varphi(x)
\int_\Gamma dk~\psi(k)e^{ikx} \nonumber
\end{eqnarray}
So that:
\begin{equation}
\hat\psi(\hat\varphi)= 
\int_{-\infty}^{+\infty}dx~\hat\psi(x)\hat\varphi(x)
\label{e12}
\end{equation}
where
\begin{equation}
\hat{\psi}(x)=\int_\Gamma
dk~\psi(k)e^{ikx} ~.
\label{e13}
\end{equation}
$\psi(k)$ is a density for $\hat\varphi(x)$ ($\in \hat{\zeta}$). 
Of course, if $\psi(k)=a(k)$ (entire analytic), $\hat\psi(x)=0$. So
that $\hat\psi(x)$ is not altered if we add $a(k)$ to $\psi(k)$. 

It should be clear that Eq.\ (\ref{e13}) is to be understood in the
sense of distributions. For example, if $\psi(k)$ is the density for
the unit functional (Eq.\ (\ref{e4})): 
\begin{eqnarray}
\hat\psi(x)&=&\frac{1}{2}\int_\Gamma dk~Sg(Im\;k)
e^{ikx} \nonumber \\
&=&\frac{1}{2}\int_{-\infty}^{+\infty} dk~ e^{ikx}-
\frac{1}{2}\int_{+\infty}^{-\infty} dk~ e^{ikx} \nonumber \\
\hat\psi(x)&=&2\pi\delta(x) 
\label{e14}
\end{eqnarray}

\section{Pseudo-differential Operators \label{s3}}
Our aim is to work properly with some non-local  pseudo-differential
operators. For this reason we are going to introduce the following
operation on the  functionals $\hat\psi$. 

Let us consider   a function  $f(k)$ such that: $f(k)$ is analytic in 
$\{k\in C/ |Im\;k|>\beta\}$ and $f(k)/k^\beta$ is bounded continuous
in  $\{k\in C/ |Im\;k|\ge\beta\}$, $\beta$ depends on $f(k)$,
$\beta\in N$.

Then we define
\begin{equation}
f\hat\psi(\hat\varphi)=\int_{-\infty}^{+\infty}
\left[ f\left(-i\frac{d~}{dx}\right)\hat\psi(x)\right]\hat\varphi(x)
\label{e15}
\end{equation}
where:
\begin{equation}
f(-i\frac{d~}{dx})\hat\psi(x)=\int_\Gamma dk~f(k)\psi(k)e^{ikx}
\label{e16}
\end{equation}
and where the path $\Gamma$ runs from $-\infty$ to $+\infty$
 along $Im\;k>\beta+\rho$ and  from $+\infty$ to $-\infty$
 along $Im\;k<-\beta-\rho$. 

We know that the functional $\hat\psi(\hat\varphi)$ does not change
when we add an arbitrary entire function $a(k)$ to $\psi(k)$.
However, such  an addition in Eq.\ (\ref{e16}) gives rise to a new
term 
\begin{equation}
A(x)=\int_\Gamma dk~f(k)a(k) e^{ikx}
\label{e17}
\end{equation}
When $f(z)$ is an entire function, $A(x)\equiv 0$. 
For example for polynomial functions of the derivative operator.  
If $f(z)=z^{-1}$, $f(-i\frac{d~}{dx})$ is the inverse of the derivative
(an integration). Equation (\ref{e16}) gives a primitive of
$\hat\psi(x)$. 
The additional term (\ref{e17}) is:
\begin{equation}
A(x)=\int_\Gamma dk~ \frac{a(k)}{k} e^{ikx}=\int_L dk~ \frac{a(k)}{k}
e^{ikx}
\label{e18}
\end{equation}
where $L$ is a loop around the origin. So that: 
\[
A(x)=-2\pi i a(0)=\mbox{arbitrary constant}
\]
of course, an integration should give a primitive plus an arbitrary
constant.

Analogously, a double (iterated) integration, with $f(z)=z^{-2}$,
gives a primitive plus (\ref{e17}): 
\[
A(x)=\int_L dk~ \frac{a(k)}{k^2} e^{ikx}= \gamma+\delta x
\]
where $\gamma$ and $\delta$ are arbitrary constants, as it should be. 

It is understandable that a more complex structure of singularities of
$f(z)$ gives rise to a more complicated $A(x)$. 

To solve some pseudo-differential equations, we can work directly with
the densities, the null functional having an arbitrary entire function
as density. (an arbitrary constant has $k^{-1}a(k)$ as density, etc.)

For a solution to a homogeneous equation:
\begin{equation}
f(-i\frac{d~}{dx})\hat\psi(x)=0
\label{e19}
\end{equation}
we write
\[
f(k)\psi(k)=a(k)
\]
whose solution is:
\begin{equation}
\psi(k)=f^{-1}(k)a(k)
\label{e20}
\end{equation}
then replacing in  Eq.\ (\ref{e13}) we obtain:
\begin{equation}
\hat{\psi}(x)=\int_\Gamma
dk~f^{-1}(k)a(k)e^{ikx}
\label{hom}
\end{equation}
In the case where the singularities of $f^{-1}$ are concentrated on the real 
axis, the path $\Gamma$ can be deformed to get:
\begin{equation}
\hat{\psi}(x)=\int_{-\infty}^{+\infty} dk~
\left[ f^{-1}(k+i 0)-f^{-1}(k-i 0)\right] e^{ikx} 
\label{hom'}
\end{equation}

To solve the inhomogeneous equation:
\begin{equation}
f(-i\frac{d~}{dx})\hat\psi(x)=\hat\chi(x)
\label{e21}
\end{equation}
where $\hat\chi(x)$ is a given (generalized) function, we write
\begin{eqnarray}
f(k)\psi(k)&=& \chi(k)+ a(k) \nonumber \\
\psi(k)&=&f^{-1}(k)\chi(k)+f^{-1}(k)a(k)  \label{e22}
\end{eqnarray}
where $\hat{\chi}(x)=\int_\Gamma
dk~(\chi(k)+a(k))\exp{ikx}=\int_\Gamma
dk~\chi(k)\exp{ikx}$.
The last term in Eq. (\ref{e22}) can be recognized as a general solution to the homogeneous equation (\ref{e19}). 

If in (\ref{e21}) we choose  $\hat\chi(x)=\delta(x)$, we get the
equation for the Green function:
\begin{equation}
f(-i\frac{d~}{dx})\hat{G}(x)=\delta(x)
\label{e23} 
\end{equation}
As a density for Dirac's function is $\frac{1}{2}Sg(Im\;k)$
(cf.\ (\ref{e14})), we obtain from (\ref{e22}):
\begin{equation}
G(k)=\frac{1}{2}f^{-1}(k) Sg(Im\;k)+f^{-1}(k)a(k)
\label{e24a}
\end{equation}
Eq.\ (\ref{e24a}) shows that, as usual, the Green function is defined up to a solution to the homogeneous equation. The different choices of $a(k)$ determine the different prescriptions (advanced, retarded, Feynman, Wheeler, etc.) to avoid the singularities, when evaluating the
Fourier transform. In the case where $a=0$ we have:
\begin{equation}
G(k)=\frac{1}{2}f^{-1}(k) Sg(Im\;k)+f^{-1}(k)a(k)
\label{e24}
\end{equation}
It is interesting to find the Fourier transform of (\ref{e24}):
\[
\hat{G}(x)=\frac{1}{2}\int_\Gamma dk~ f^{-1}(k) Sg(Im\;k)e^{ikx}
\]      
I.\ e.\ :
\begin{eqnarray}
\hat{G}(x)&=&\frac{1}{2}\int_{-\infty}^{+\infty} dk~
\left( f^{-1}(k+i 0)+f^{-1}(k-i 0)\right) e^{ikx} \nonumber \\
\hat{G}(x)&=&g_+(x)+g_-(x)
\label{e25}
\end{eqnarray}
where 
\begin{equation}
g_{\pm}=\frac{1}{2}\int_{-\infty}^{+\infty} dk~
 f^{-1}(k\pm i 0) e^{ikx}
\label{e26}
\end{equation}

Let us now consider the composition of two operators $f_1$ and
$f_2$,  to give $f(-i\frac{d~}{dx})=f_1(-i\frac{d~}{dx})\circ  f_2(-i\frac{d~}{dx})$.
For the densities we have $f(k)=f_1(k)f_2(k)$ and practically nothing
changes in the analysis we have made above. However, if we apply the
two operators in succession we have the following situation:  

For the equation
\begin{equation}
f_2(-i\frac{d~}{dx})\left[ f_1(-i\frac{d~}{dx})
\hat\psi \right]=0 \label{e27}
\end{equation}
the square bracket is a solution of the homogeneous equation for
$f_2$. Accordingly:
\[
f_1(-i\frac{d~}{dx})\hat\psi(x)=\int_\Gamma dk~f_2^{-1}a(k)e^{ikx}
\]
Now we have an inhomogeneous equation for $f_1$, whose solution is
\begin{equation}
\hat\psi(x)=\int_\Gamma dk~f_1^{-1}(k)f_2^{-1}(k)a(k)e^{ikx}+
\int_\Gamma dk~f_1^{-1}(k)b(k)e^{ikx}
\label{e28}
\end{equation}
If we interchange $f_1$ and $f_2$ in Eq.\ (\ref{e27}), the solution
(\ref{e28}) changes is an obvious way. 

On the other hand, if $f_1$  and $f_2$  are first composed to give
$f=f_1\circ f_2$, the corresponding solution (\ref{e28}) loses its
last term. 
 
\section{Some Examples \label{s4}}
\subsection{Rational Functions}
We consider first  a rational function (see also Ref.\ \cite{18},
Ch.\ 2, $\S 1.5$), 
\begin{equation}
f(-i\frac{d~}{dx})=\frac{P(-i\frac{d~}{dx})}{Q(-i\frac{d~}{dx})}
\label{e29}
\end{equation}
\begin{equation}
P(k)=\prod_{i=1}^{n}(k-k_i)~~~;~~~Q(k)=\prod_{j=1}^{n}(k-k_j')
\label{e30}
\end{equation}
\[
(k_i\neq k_j' ~\mbox{for all}~i~\mbox{and}~j)
\]

The solution to the homogeneous equation:
\begin{equation}
\frac{P(-i\frac{d~}{dx})}{Q(-i\frac{d~}{dx})}\hat\psi(x)=0
\label{e31}
\end{equation}
is given by (\ref{e19}) and (\ref{e20})
\begin{equation}
\hat\psi(x)=\int_\Gamma dk~ \frac{Q(k)}{P(k)}a(k)e^{ikx}
\label{e32}
\end{equation}
If we rewrite (\ref{e31}) in the form: 
\begin{equation}
Q^{-1}(-i\frac{d~}{dx})\left[ P(-i\frac{d~}{dx})\hat\psi(x)\right]=0
\label{e33}
\end{equation}
The solution (\ref{e28}) is:
\begin{eqnarray}
\hat\psi(x)&=&\int_\Gamma dk~ 
\frac{Q(k)}{P(k)} a(k)e^{ikx}  +
\int_\Gamma dk~ 
\frac{b(k)}{P(k)} e^{ikx}         \nonumber \\
\hat\psi(x)&=&\int_\Gamma dk~ 
\frac{1}{P(k)}\left[ Q(k)a(k)+b(k)\right] e^{ikx}
\label{e34}
\end{eqnarray}
 
The square bracket in (\ref{e34}) is another arbitrary entire
function. So that (\ref{e34}) is in fact equivalent to (\ref{e32}). 

If now we write the equation:
\begin{equation}
P(-i\frac{d~}{dx})\left[ Q^{-1}(-i\frac{d~}{dx})\hat\psi(x)\right]=0
\label{e35}
\end{equation}
The solution (\ref{e28}) is:
\[
\hat\psi(x)=\int_\Gamma dk~ 
\frac{Q(k)}{P(k)} a(k)e^{ikx}  +
\int_\Gamma dk~ 
Q(k )b(k) e^{ikx}    
\]
But the last integral is zero, as the product $Q b \exp{ikx}$ is entire
analytic.

The three forms (\ref{e31}), (\ref{e32}) and (\ref{e33}) are completely
equivalent. The order of the operators $P$ and $Q$ is inmaterial. 

For the actual solution of (\ref{e32}), we take into account that 
\begin{equation}
\frac{1}{P(k)}=\frac{1}{\prod (k-k_i)}=\sum_{i=1}^{n}
\frac{\alpha_i}{k-k_i}
\label{e36}
\end{equation}
where $\alpha_i$ are appropriate constants (For the sake of simplicity
we have assumed that all roots are simple). 
 
The only singularities of the integrand in (\ref{e32}) are the poles
at  $k=k_i$. The integral can then be evaluated  by Cauchy's theorem:

\begin{eqnarray}
\hat\psi(x)&=& \sum_{i=1}^{n}\alpha_i \int_\Gamma dk~
\frac{Q(k_i)}{k-k_i}a(k)e^{ikx} \nonumber \\
&=& 2\pi i \sum_{i=1}^{n}\alpha_i Q(k_i)a(k_i)e^{ik_ix}\nonumber \\
&=& \sum_{i=1}^{n}\beta_i e^{ik_ix}   
\label{e37}
\end{eqnarray}
where $\beta_i$ are arbitrary constants.  It is easy to check that
(\ref{e37}) solves equations (\ref{e31}), (\ref{e33}) and (\ref{e35}).

\subsection{Power Functions}
For another (less simple) example, let us take the power function:
\begin{equation}
f(z)=z^\alpha
\label{e38}
\end{equation}
where we introduce a cut along the negative real axis. 

For the homogeneous equation
\begin{equation}
\left(-i\frac{d~}{dx}\right)^{\alpha}\hat\psi(x)=0
\label{e39}
\end{equation}
we have: $\psi(k)=k^{-\alpha}a(k)$ and 
\begin{equation}
\hat\psi(x)=\int_\Gamma dk~ k^{-\alpha}a(k)e^{ikx}
\label{e40}
\end{equation}
The integrand is analytic, except on the (negative) real axis.
We may write: 
\begin{equation}
\hat\psi(x)=\int_{-\infty}^{\infty} dk~ 
\left[ (k+i 0)^{-\alpha}-(k-i 0)^{-\alpha}\right] a(k)e^{ikx}
\label{e41}
\end{equation}
But, (Ref.\ \cite{18}, Ch.\ 1, $\S 3.6$)
\begin{eqnarray}
(k+i 0)^{-\alpha}&=&k_+^{-\alpha}+e^{-ik\alpha} k_-^{-\alpha}
\nonumber\\
(k-i 0)^{-\alpha}&=&k_+^{-\alpha}+e^{ik\alpha} k_-^{-\alpha}
\label{e42}
\end{eqnarray}
So that:
\begin{equation}
\hat\psi(x)=-2i \sin{(\pi \alpha)}
\int_{-\infty}^{+\infty} dk~ k_-^{-\alpha} a(k) e^{ikx}
\label{e43}
\end{equation}

The equation for a Green function is: 

\[
\left(-i\frac{d~}{dx}\right)^{\alpha}\hat{G}(x)=\delta(x)
\]
I.\  e.: (cf.\ Eq.\ (\ref{e4})):
\begin{eqnarray}
\hat{G}(x)&=&\frac{1}{2}\int_\Gamma dk~ k^{-\alpha}Sg(Im\;k)
e^{ikx}
\nonumber \\
\hat{G}(x)&=&\frac{1}{2}\int_{-\infty}^{+\infty}dk~ 
\left[ (k+i0)^{-\alpha}+(k-i0)^{-\alpha} \right] e^{ikx}
\label{e44}
\end{eqnarray}
For the Fourier transform we use Ref.\ \cite{18} (Ch.\ 2. $\S 2.3$)
\begin{equation}
{\em F}(k\pm i 0)^{-\alpha}=\frac{e^{\mp i\frac{\pi}{2}\alpha}}
{\Gamma(\alpha)}   x_{\mp}^{\alpha-1}
\label{e45}
\end{equation}
Then:
\begin{equation}
\hat{G}(x)=\frac{1}{2\Gamma(\alpha)}\left[ e^{i\frac{\pi}{2}\alpha}
x_+^{\alpha-1}+e^{-i\frac{\pi}{2}\alpha}
x_-^{\alpha-1}\right]
\label{e46}
\end{equation}

To see the connection  with the fractional derivative, we choose to add
to (\ref{e44}),  the solution (\ref{e41}) to the homogeneous equation
with $a(k)\equiv -1/2$. The new Green-function is the result of
(\ref{e44}) minus one half of (\ref{e41}):
\begin{equation}
\hat{G}_+(x)=\int_{-\infty}^{+\infty} dk~ (k- i0)^{-\alpha} e^{ikx}
= \frac{e^{i\frac{\pi}{2}\alpha}}{\Gamma(\alpha)}
x_+^{\alpha-1}
\label{e47}
\end{equation}
Except  of course for the exponential factor, Eq.\ (\ref{e47})
coincides with the fractional derivative considered in Ref.\
\cite{18}(Ch.\ 1, $\S 5.5$). 

If instead of adding (\ref{e41}) with $a\equiv -1/2$, we add the same
function but with $a\equiv +1/2$ we get:

\begin{equation}
\hat{G}_-(x)= \frac{e^{-i\frac{\pi}{2}\alpha}}{\Gamma(\alpha)}
x_-^{\alpha-1}
\label{e48}
\end{equation}
which is also a fractional derivative operator. 

\section{Functions of the D'Alambertian \label{s5}}
In this section we are going to examine some non-local  functions $f(\Box)$
of $\Box=\partial_0^2-\vec{\partial}^2$, where $\vec{\partial}^2$ is
the  ($d-1$)-dimensional laplacian operator. 

For $f(z)$ we will take an analytic function such as $z^{\alpha}$ or
$(z+m^2)^\alpha$, with a cut along the negative real axis, running from
$-\infty$ to zero or $-m^2$, respectively. 

The ultra distributions (Eq.\ (\ref{e2})) depend now on an
($n-1$)-dimensional vector $\vec{k}$ as parameter
$\psi(k_0)\rightarrow \psi(k_0,\vec{k})$. 

The Fourier transform of Eq.\ (\ref{e13}) gives:

\begin{equation}
\hat\psi(k_0,\vec{k})=\int_\Gamma dk_0~\psi(k_0,\vec{k})e^{ik_0x_0}
\label{e49}
\end{equation}
And, with the usual Fourier transform in the space of the parameters
$\vec{k}$, we obtain:

\begin{equation}
\hat\psi(x)=\hat\psi(x_0,\vec{x}=\int_{-\infty}^{+\infty}
d\vec{k}~ \hat\psi(x_0,\vec{k})e^{i\vec{k}\vec{x}}
\label{e50}
\end{equation}

We define the operation $f(\Box)$ on $\hat\psi(x)$
by:
\begin{equation}
f(\Box)\hat\psi(x)=\int d\vec{k}~ f(\partial_0^2+\vec{k}^2)
\hat\psi(x_0,\vec{k})e^{-i\vec{k}\vec{x}}
\label{e51}
\end{equation}
And (Cf.\ Eq.\ (\ref{e16})):
\begin{equation}
f(\partial_0^2+\vec{k}^2)\hat\psi(x_0,\vec{k})=
\int_\Gamma  dk_0~ f(-k_0+\vec{k}^2)\psi(k_0,\vec{k})e^{+ik_0x_0}
\label{e52}
\end{equation}

We can now solve the homogeneous (n-dimensional) equation:
\begin{equation}
f(\Box)\hat\psi(x)=0
\label{e53}
\end{equation}
According to (\ref{e51}), Eq.\ (\ref{e53}) implies:
\[
f(\partial_0^2+\vec{k}^2)\hat\psi(x_0,\vec{k})=0
\]
And, due to (\ref{e52}), we have for the densities:
\[
 f(-k_0^2+\vec{k}^2)\psi(k_0,\vec{k})=a(k_0^2+\vec{k}^2)
\]
\begin{equation}
\psi(k_0,\vec{k})=f^{-1}(-k_0^2+\vec{k}^2)a(k_0^2+\vec{k}^2)
\label{e54}
\end{equation}
Where $a(k_0^2+\vec{k}^2)$ is an arbitrary entire function of $k_0$,
for any value of  $\vec{k}$. 
 
>From (\ref{e49}) and (\ref{e54}) we get:
\begin{equation}
\hat\psi(x_0,\vec{k})=\int_\Gamma dk_0~f^{-1}(-k_0^2+\vec{k}^2)
a(k_0,\vec{k})e^{ik_0x_0}
\label{e55}
\end{equation}
The analytic function of $k_0$, $f^{-1}(-k_0^2+\vec{k}^2)$ presents a
cut along the real axis, running from $-\infty$ to $k_0=-\omega$,
another twin cut from $k_0=+\omega$ to 
$+\infty$ ($\omega=+\sqrt{\vec{k}^2+m^2}$). 

The integration in (\ref{e55}) can then be taken along the real axis:
\begin{equation}
\hat\psi(x_0,\vec{x})=\int_{-\infty}^{+\infty} dk_0~
\Delta(f^{-1}) a(k_0,\vec{k}) e^{ik_0x_0}
\label{e56}
\end{equation}
where $\Delta$ is the discontinuity at the cuts:
\begin{equation}
\Delta(f^{-1})=f^{-1}(-(k_0+i 0)^2+\vec{k}^2)-
f^{-1}(-(k_0-i 0)^2+\vec{k}^2)
\label{e57}
\end{equation}
\[
\left\{ \Delta(f^{-1})=0,~\mbox{for}~-\omega\le k_0\le\omega~,~\omega
=+\sqrt{\vec{k}^2} ~
\mbox{when}~ m=0 \right\}
\]

For $f=(-k_0^2+\vec{k}^2+m^2)^\alpha=(-k_0^2+\omega^2)^\alpha$, 
we have:
\begin{eqnarray}
\Delta(f^{-1})&=&(-k_0^2+\omega^2-i 0 Sg(k_0))^{-\alpha}-
(-k_0^2+\omega^2+i 0 Sg(k_0))^{-\alpha}\nonumber\\
&=&Sg(k_0)\left[(-k_0^2+\omega^2-i 0)^{-\alpha}-
(-k_0^2+\omega^2+i 0)^{-\alpha} \right] \nonumber
\end{eqnarray}
And using (\ref{e42}):
\begin{equation}
\Delta(f^{-1})=
2 i \sin{(\pi\alpha)}Sg k_0(-k_0^2+\omega^2)^{-\alpha}_-=
2 i \sin{(\pi\alpha)}Sgk_0(k_0^2-\omega^2)_+^{-\alpha}
\label{e58}
\end{equation}

For $\alpha=1$, Eq.\ (\ref{e53}) is Klein-Gordon equation. The weight
function given by (\ref{e58}), has a pole for this value of $\alpha$
(Ref.\ \cite{18}, Ch.\ 3, $\S 3.4$).
The residue is $\delta(k_0^2-\omega^2)=\delta(k^2-m^2)$. 
For $\alpha\rightarrow 1 $, Eq.\ (\ref{e58}) gives 
$\Delta\rightarrow \delta(k^2-m^2)$, which is the well-known
invariant free wave solution. For a fractional
$\alpha$, the weight function gives a continuum with 
$k^2\ge m^2$. In this sense we can say that (\ref{e58}) represents
free waves corresponding to a continuum of masses $k^2=\mu^2\ge m^2$.
The free wave $\delta(k^2-m^2)$ concentrated on $k^2=m^2$, changes for
fractional $\alpha$ into $(k^2-m^2)_+^{-\alpha}$, which is spread from
$k^2=m^2$ to $k^2\rightarrow\infty$ (see also Ref.\ \cite{20}). 

For the fractional wave equation:
\begin{equation}
\Box^\alpha\hat\psi(x)=0
\label{e59}
\end{equation}
we can repeat the analysis made above. In this case:
\[
f=(-k_0^2+\vec{k}^2)^\alpha
\]
\begin{eqnarray}
\Delta(f^{-1})&=&(-k_0^2+\vec{k}^2-i 0 Sg(k_0))^{-\alpha}-
(-k_0^2+\vec{k}^2+i 0 Sg(k_0))^{-\alpha}\nonumber\\
&=&Sg(k_0)\left[(-k_0^2-i 0)^{-\alpha}-
(-k_0^2+i 0)^{-\alpha} \right] \nonumber
\end{eqnarray}
\begin{equation}
 \Delta(f^{-1})=2 i \sin{(\pi\alpha)}Sgk_0(k_0^2)_+^{-\alpha}
\label{e60}
\end{equation}
Again, for $\alpha\rightarrow 1$, (\ref{e59}) is the usual wave
equation and $\Delta(f^{-1})\rightarrow \delta(k^2)$, 
which is the elementary solution with support on the surface of the
light-cone. When $\alpha$ is not an integer, 
Eq.\ (\ref{e60}) implies that we can find free wave components for
any positive mass squared. 

Let us now find a Green function for the fractional wave equation:
\begin{equation}
\Box^\alpha\hat{G}(x)=\delta(x)
\label{e61}
\end{equation}
\[
\Box^\alpha\hat{G}(x)=\int d\vec{k}~ (\partial_0^2+\vec{k^2})^\alpha
\hat{G}(x_0,\vec{k}) e^{-i\vec{k}\vec{x}}=
\int d\vec{k}~\delta(x_0)e^{-i\vec{k}\vec{x}}
\]
\[
(\partial_0^2+\vec{k}^2)^\alpha
\hat{G}(x_0,\vec{k})=\delta(x_0)
\]
And we have for the densities:
\[
(-k_0^2+\vec{k}^2)^\alpha G(k_0,\vec{k})=\frac{1}{2}Sg Im k_0
\]
\[
G(k_0,\vec{k})=\frac{1}{2}(-k_0^2+\vec{k}^2)^{-\alpha}
 Sg Im k_0
\]
\[
\hat{G}(x_0,\vec{k})=\frac{1}{2}\int_{-\infty}^{+\infty} dk_0~
(-k_0^2+\vec{k}^2)^{-\alpha} Sg Im k_0 e^{ik_0x_0}
\]
So that:
\begin{equation}
\hat{G}(x_0,\vec{k})=\frac{1}{2}\int_{-\infty}^{+\infty} dk_0~
\left[(-k^2-i 0)^{-\alpha}+(-k^2+i 0)^{-\alpha}\right]
e^{ik_0x_0}
\label{e62}
\end{equation}
The integrand in (\ref{e62}) does not depend on the sign of $k_0$.
This Wheeler Green function for the fractional wave equation
(\ref{e61}) is an extension (for space-time) of the definition given
in eq.\ (\ref{e9}). 

The two terms in the square bracket of (\ref{e62}) are, 
respectively, the causal and anticausal Green functions
for the fractional wave equation. For $\alpha=1$ they are the usual
Feynman propagator and its complex conjugate. 
The Fourier transforms of those two terms can be found in Ref.\
\cite{18}, Ch 3, $\S 2.6$. 
Using (\ref{e42}) we can write the square brackets as :
\[
[~]=(k^2)_-^{-\alpha}+e^{i\pi\alpha}(k^2)_+^{-\alpha} 
(k^2)_-^{-\alpha}+e^{-i\pi\alpha}(k^2)_+^{-\alpha}
\]
so, 
\begin{equation}
\frac{1}{2}\left[(-k^2-i 0)^{-\alpha}+(-k^2+i 0)^{-\alpha}\right]=
(k^2)_-^{-\alpha}+\cos{\pi\alpha} (k^2)_+^{-\alpha}
\label{e63}
\end{equation}

Other Green functions can be  found by adding to (\ref{e62})
solutions to the homogeneous equation. 
Adding together (\ref{e56}) (with (\ref{e58}) and $a=\pm1/2$) and
(\ref{e62}) (with (\ref{e63})), we get:
\begin{equation}
\hat{G}_{\pm}(x_0,\vec{k})=
\int_{-\infty}^{^\infty} dk_0~
\left[ (k^2)_-^{-\alpha}+e^{\pm i\pi\alpha Sg k_0}
(k^2)_+^{-\alpha}\right]
e^{ik_0x_0}
\label{e64}
\end{equation}
The two terms of (\ref{e62}) and the two Green functions of 
(\ref{e64}), coincide with the four Lorentz-invariant propagators for
the fractional wave equation found in Ref.\  \cite{11}.

It is possible to repeat this procedure to find Green functions for
the fractional Klein-Gordon equation:
\begin{equation}
(\Box+m^2)^\alpha\hat{G}(x)=\delta(x)
\label{e65}
\end{equation}
Instead of (\ref{e62}), we now have:
\begin{equation}
\hat{G}(x_0,\vec{k})=\frac{1}{2}\int_{-\infty}^{+\infty} dk_0~
\left[(-k^2+m^2-i 0)^{-\alpha}+(-k^2+m^2+i 0)^{-\alpha}\right]
e^{ik_0x_0}
\label{e66}
\end{equation}
In this way we obtain the causal and anticausal propagators for the
fractional Klein-Gordon equation. The Feynman function coincides with
the propagator used in Ref.\ \cite{7} to regularize the matrix elements
of Quantum Electrodynamics. 

\section{Discussion \label{s6}}

The space $\zeta$ of ultra analytic functions, defined in section
\ref{s2}, seems to be an appropriate base of test functions for the
development of a framework adapted to the needs of those who want to
work with non-local  pseudo-differential operators. Its dual space $\zeta'$
contains the propagators appearing in perturbative Quantum Field
Theories. It is also flexible enough to allow for them a
representation in terms of ultra distributions which is both, simple and
general. 

The Fourier transformed spaces $\hat{\zeta}$ and $\hat{\zeta}'$,
translate the functions of $\zeta$ and  $\zeta'$ into functions of
coordinates where the pseudo-differential operators are supposed to
act.  In this way we can define non-local  operations on (generalized) 
functions of the coordinates.  These operators are extensions of the
ordinary derivative operators, which are now particular cases in a
wider scheme. 

We have written and solved homogeneous and inhomogeneous equations. In
particular, we have found Green functions for some non-local operators. 

We have also shown the connection with the operators $\Box^\alpha$,
defined and discussed in Ref.\ \cite{11}, and with the fractional
derivative of Ref.\ \cite{18}. In  these references no attempt was
made to solve the respective homogeneous equations. We here exhibited
the corresponding solutions. As a consequence, we are able to show
that the free solutions of the wave equation, or the Klein-Gordon
equation, with their sharp masses, are transformed when the equations
are fractional, into a superposition of a continuum of masses with
support in the interior of the light cone. Furthermore, the causal
propagators for the fractional equations, are seen to coincide with
the analytically regularized propagators introduced in Ref.\ \cite{7}.
It is then possible to  interpret this regularization method, as the
matrix elements one would write for a non-local  theory having no
ultraviolet divergences.  The usual infinities appear as poles for the
local limit $\alpha\rightarrow 1$. 

We see then that with the chosen procedures we have convenient tools
with which we can handle different pseudo-differential non-local  problems.
The scheme also provides a base for the quantization of non-local  field
theories \cite{20}.
 
\section*{Acknowledgements}
This work was partially supported by Centro Latinoameriacano de F\'\i sica (CLAF),
Conselho Nacional de Desenvolvimento Cient\'\i fico e Tecnol\'ogico (CNPq, Brasil), Consejo Nacional de Investigaciones Cient\'{\i}ficas y T\'ecnicas'' (CONICET, Argentina) and
Comisi\'on de Investigaciones Cient\'{\i}ficas de la Pcia.\ de Bs.\  As.\ (CIC, Argentina).


\newpage
\begin{thebibliography}{99}

\bibitem{1} J.\ A.\ Wheeler and R.\ P.\ Feynman; 
Rev.\ Mod.\ Phys.\ {\bf 17} (1945), 157. 

\bibitem{3}M.\ N.\ Volkov and G.\ V.\ Efimov;
Sov.\ Phys.\ Usp.\ 23 (1980), 94. 

\bibitem{4}N.\ V.\ Krasnokov; Theor.\ Math.\ Phys.\ 
{\bf 73} (1987), 1184. 

\bibitem{5}D.\ A.\ Eliezer and R.\ P.\ Woodard; 
Nucl.\ Phys.\ {\bf B325} (1984), 389. 

\bibitem{6}A.\ Hata; Nucl.\ Phys.\ {\bf B329} (1990), 698. 

\bibitem{2}A.\ Pais and G.\ E.\ Uhlembeck; Phys.\ Rev.\ 
{\bf 79} (1950), 145. 

\bibitem{7}C.\ G.\ Bollini, J.\ J.\ Giambiagi and 
A.\ Gonz\'ales Dom\'\i ngues; Il Nuovo Cimento {\bf 31} (1964),  550.  

\bibitem{8}A.\ O.\ Barvinsky and C.\ A.\ Vilkovisky; 
Nucl.\ Phys.\ {\bf B282} (1987), 163. 

\bibitem{9}A.\ O.\ Barvinsky and C.\ A.\ Vilkovisky; 
Nucl.\ Phys.\ {\bf B333} (1990), 471. 

\bibitem{10} D.\ A.\ R.\ Dalvit and F.\ D.\ Mazzitelli;
Phys.\ Rev.\ {\bf D50} (1994), 1002. 

\bibitem{11}C.\ G.\ Bollini and J.\ J.\ Giambiagi;
Jour.\ of Math.\ Phys.\  {\bf 34} (1993), 610. 

\bibitem{12} J.\ J.\ Giambiagi; ``Huygens Principle in $2N+1$
Dimensions for non-local operators $\Box^\alpha$'';
(preprint) CBPF-NF-025/91-Brasil.

\bibitem{13}R.\ L.\ P.\ G.\ Amaral and E.\ C.\ Marino;
J.\ Phys.\ {\bf A25} (1992), 5183. 

\bibitem{14}E.\ C.\ Marino; Phys.\ Lett.\ {\bf 263B} (1991), 63. 

\bibitem{14b}D.\ G.\ Barci, C.\ D.\ Fosco and L.\ E.\ Oxman;``On
Bozonization in $3$ dimensions'',  to be published in Phys.\ Lett.\ {\bf B} (1996). 

\bibitem{15}E.\ C.\ Marino; Nucl.\ Phys.\ {\bf B408}[FS] 
(1993), 551. 

\bibitem{ffp}G.\ M.\ Zaslavsky; Physica {\bf D76}  (1994)   110. 

\bibitem{fd1}W.\ Wyss; J.\ Math.\ Phys.\ {\bf 27} (11) (1986), 2782.  
\bibitem{fd2} W.\ R.\ Schneider and W.\ Wyss; J.\ Math.\ Phys.\ {\bf 30} (1) (1989), 134.   

\bibitem{20}D.\ G.\ Barci, L.\ E.\ Oxman and M.\ C.\ Rocca; 
``Canonical Quantization of Non-Local Field Equations'',
to be published in Int.\ Jour.\ Mod.\ Phys.\ {\bf A}. 

\bibitem{16}L.\ Sebasti\~ao e Silva; Math.\ Annalen Bd {\bf 136}
(1958), 58. 

\bibitem{17}M.\ Hasuni; T\^ohoku  Math.\ J.\ {\bf 13} (1961), 94. 

\bibitem{18}I.\ M.\ Gelfand and  G.\ E.\ Shilov;
``Les Distributions'', Tome I, Dunod, Paris, 1972.  

\bibitem{19}C.\ G.\ Bollini, L.\ E.\ Oxman and M\ Rocca; 
J.\ Math.\ Phys.\ {\bf 35} (1994), 4429. 

\end{thebibliography}
\end{document}